\documentclass{aa}
\usepackage{graphicx}
\usepackage[varg]{txfonts}
\usepackage{longtable}
\usepackage{lscape}

\begin{document}

 \title{Solar chromospheric emission and magnetic structures from plages to intranetwork: Contribution of the very quiet Sun}

   \titlerunning{Solar chromospheric emission and magnetic structures from plages to intranetwork}

   \author{N. Meunier \inst{1}   
          }
   \authorrunning{Meunier et al.}

   \institute{
Univ. Grenoble Alpes, CNRS, IPAG, F-38000 Grenoble, France\\
  \email{nadege.meunier@univ-grenoble-alpes.fr}
             }

\offprints{N. Meunier}

   \date{Received ; Accepted}

\abstract
{ We need to establish a correspondence between the magnetic structures generated by models and usual stellar activity indexes to simulate radial velocity time series for stars less active than the
Sun. This is necessary to compare the outputs of such models with observed radial velocity jitters and is critical to better understand the impact of stellar activity on exoplanet detectability. }
{We propose a coherent picture to describe the relationship between magnetic activity, including the so-called quiet Sun regions, and the chromospheric emission using the Sun as a test-bench and a reference.}
{We analyzed a long time series of Michelson Doppler imaging (MDI) magnetograms jointly with chromospheric emission time series obtained at Sacramento Peak and Kitt Peak observatories. This has allowed us to study the variability in the quiet Sun over the solar cycle, and then, based on available relations between magnetic fields in active structures and chromospheric emission, to propose an empirical reconstruction of the solar chromospheric emission based on all contributions. }
{We show that the magnetic flux covering the solar surface, including in the quieted regions, varies in phase with the solar cycle, suggesting a long-term relationship between the global dynamo and the contribution of all components of solar activity. We have been able to propose a reconstruction of the solar S-index, including a relationship between the weak field component and its chomospheric emission, which is in good agreement with the literature. This allows us to explain that stars with a low average chromospheric emission level exhibit a low variability.}
{We conclude that weak flux regions significantly contribute to the chromospheric emission; these regions should be critical in explaining the lower variability associated with the lower average activity level in other stars as compared to the Sun and estimated from their chromospheric emission. }


\keywords{Sun: activity -- Sun: chromosphere -- Sun: faculae, plages -- Sun: magnetic fields -- Stars: activity  -- Stars: solar-type} 

\maketitle

\section{Introduction}

Stellar activity has a critical impact on the detectability of low-mass exoplanets around solar-type stars. In this context, using the Sun as a reference and a test bench is extremely useful. After modeling solar radial velocities (RV) over a solar cycle from observed solar features, spots, and plages \cite[][]{lagrange10,meunier10a}, we generated solar structures using a model \cite[][]{borgniet15} to be able to produce realistic and complex RV time series for stars other than the Sun. To be able to compare the results with observations, it is important to understand how to relate the activity level of a star defined by its magnetic structures (as produced by such a simulation, i.e., active regions and the magnetic network) to a typical chromospheric emission index, such as the LogR'$_{\rm HK}$ or the Mount Wilson S-index, available for many stars. 

The contribution of active regions and the magnetic network to the chromospheric emission could in principle be deduced from their filling factor by using a law such as that derived by \cite{shapiro14} because there is a good correlation between the two.
However, there are a number of indications that stars with an average activity level lower than the solar one also exhibit cyclic activity and a rotational modulation. Although there is a large dispersion in variability for a given average activity level, lower activity stars tend to have a lower variability from chromospheric and photometric indexes \cite[e.g.,][]{radick98,lovis11b,shapiro13}.
\cite{schrijver89} analyzed Mount Wilson data to derive relationships between the activity level and rotation rate for stars in various B-V ranges. Their results suggest that the rotational modulation vanishes at or very near the observational lower limit in flux only, that is, for stars with an activity level much lower than the solar activity level. Similarly, \cite{garcia14} derived the rotation rates for a sample of 78 Kepler dwarf stars and found that 56\% of the stars with an activity level lower the average Sun (24\% of those with an activity level lower than the Sun at cycle minimum) have a measurable rotation rate from the modulation of the photometric light curves. In both cases the rotation rate is derived from the modulation induced by magnetic features. This suggests that these magnetic features exist at very low activity level and that their variability may mimic that of the Sun, although with a much smaller amplitude.

Therefore, although in the solar case the filling factor and the S-index are  well correlated \cite[][]{shapiro14}, there is a major difficulty in directly applying this law to stars other than the Sun, as the solar minimum (S-index above 0.16) corresponds to a filling factor close to zero, producing a S-index far above the minimum flux observed for inactive stars, for  example, 0.144 as derived by \cite{mittag13}. 

An additional component, corresponding to the magnetic fields outside active regions and magnetic network, should therefore be taken into account when modeling the S-index. 
It is indeed known that the whole solar surface is covered by magnetic fields, which are defined in the quiet Sun as intranetwork magnetic fields 
\cite[e.g.,][]{livingston71,livingston75,martin88}. Intranetwork fields have been extensively studied over the past 20 years and are typically in the 1-20 G range at the scale of a few arcseconds with a strong horizontal component. Their relationship with active regions has been debated however:\ Do they result from the global dynamo
after the decay of active regions and network or are they due to a local
dynamo?  We therefore examine the possible contribution of such magnetic fields, corresponding to the so-called quiet Sun, to the chromospheric emission.   
This is necessary to simulate stars with a lower amplitude cycle, and a S-index lower on average than the solar value at solar minimum but a filling factor of plages and network different from zero.

The quiet Sun in the following refers to weak magnetic fields outside active structures. 
This differs from the notion of basal flux, introduced by \cite{schrijver89}, which could be due to acoustic waves and could be present even without any magnetic field. In the literature, the basal flux has been indeed used with two definitions: the basal flux when no magnetic field is present, such as the minimum observed  flux of \cite{mittag13}; or the flux outside active regions and network, as studied by \cite{schrijver92} for example. In this paper, the term basal flux will refer to the first definition.
 




The objective of this paper is to propose a coherent picture of the relationship between magnetic activity (from plages to the weakest fields) and the resulting S-index in the solar case, to provide useful inputs for the modeling of other stars with a lower average activity level and lower variability. We build such a picture on constraints available in the literature and on new analysis. The outline of this paper is as follows. We first revisit the filling factor -- S-index relationship in Sect.~2 by adding a comparison with the full disk magnetic flux. In Sect.~3 we analyze the solar cycle variations. We focus our analysis  on long-term variations in the quiet Sun, and on the comparison between the activity level of solar minima (mostly the end of cycles 22 and 23). These variations provide clues for the reconstruction of a solar S-index in Sect.~4. 

\section{Relations between S-index, filling factor, and magnetic flux}

\subsection{S-index data}

In this section, we use the Ca-index that is obtained almost daily at the Sacramento Peak Observatory because significantly more points are available compared to the Kitt Peak observations. 
We first calibrated the Sacramento Peak index into a Kitt Peak Ca index as carried out by \cite{white98} and obtained the following law based on 481 common observing days:
\begin{equation}
{\rm Ca_{KP}}=-0.0034(\pm 0.0018)+1.017(\pm 0.019) {\rm Ca_{SP}}
,\end{equation}
where Ca$_{\rm SP}$ represents the Ca index from  Sacramento Peak and Ca$_{\rm KP}$ the Ca index comparable to Kitt Peak observations.
We could then use the conversion from the Kitt Peak Ca-index to the Mount Wilson S-index derived by \cite{radick98}, i.e.,
\begin{equation}
S=0.041+1.475 {\rm Ca_{KP}}
.\end{equation}
In this section we use data obtained between 1996 and 2010, which covers the temporal range for which we are able to estimate full-disk magnetic fluxes. 
Both equations 1 and 2 are used to convert the Sacramento Peak Ca-index into S. In section 3, equation 2 is also applied to Kitt Peak Ca-index observations.

\subsection{Filling factor and magnetic flux data}

In order to estimate the filling factor of active structures and the magnetic flux in various regions of the solar disk over a full solar cycle, we used full-disk magnetograms provided the Michelson Doppler imaging (MDI) on board the SOHO spacecraft \cite[][]{Smdi95}. In this paper we consider the 5 min averaged magnetograms and select one image per day when available, i.e., 4181 images obtained between 21 April 1996 and 30 December 2010. The image was selected arbitrarily at the beginning of the day, from the available images, after a quality check (i.e. ,chosen out of 1 to 15 images depending on the date).  We also performed the same analysis on 1 min magnetograms (4755 images between 21 April 1996 and 31 December 2010) to check for the robustness of the results. This gives similar results, although they are naturally shifted toward slightly larger averaged magnetic fields (because of the larger noise level). 

We applied a threshold of 100~G (significantly above the noise, see Sect.~3.1.1) on |B| maps, and identifed structures larger than four pixels (corresponding to network and plages mostly). This threshold is similar to that used in \cite{meunier10}, which allows us to establish some comparisons in future works. The analysis was made for $\mu$ (cos of the heliocentric angle) larger than 0.312 (relative distance from disk center lower than 0.95) to avoid the larger noise close the limb. Full disk values hereafter refer to this portion of the disk. An apparent filling factor ff corresponding mostly to the network and plages for each image was derived from this list of structures in ppm of the solar disk. We also computed the average |B| ($<|B|>$) over the full disk ($\mu$ also larger than 0.312). 

This joint analysis of the 5 min magetograms and Sacramento Peak indexes can be performed for 1584 days. We focused on this time series in the following subsection. 

\subsection{Analysis and interpretation}

\begin{figure} 
\includegraphics{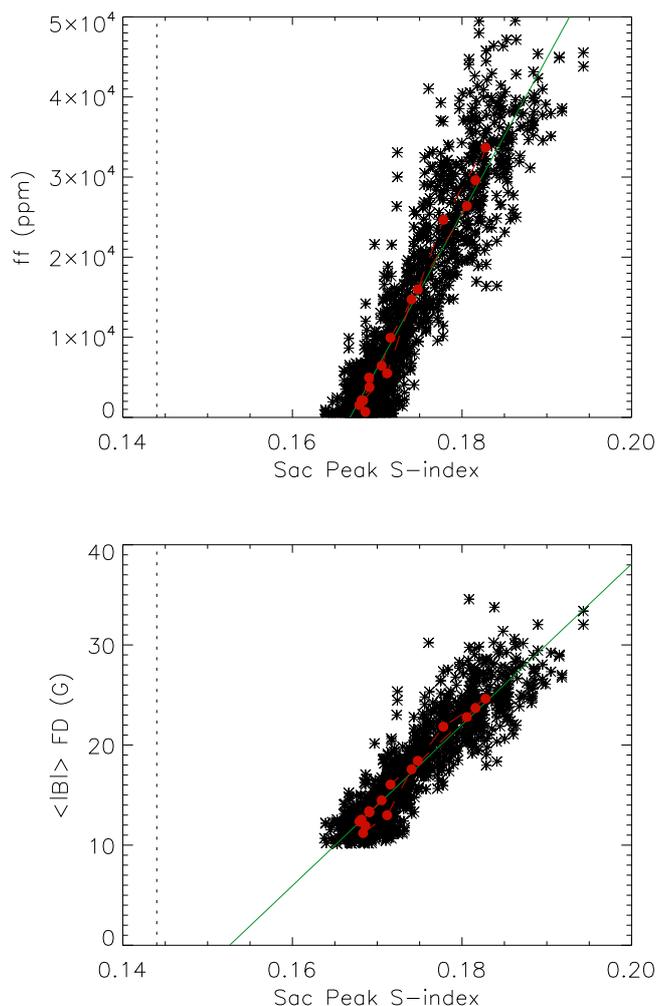}
\caption{
{\it Upper panel}:  Sacramento Peak S-index  vs. ff for all observations (black stars) after conversion into a Mount Wilson S-index. The red dots correspond to one-year binning. The green solid line is a linear fit on all points.
{\it Lower panel}: Same for $<|B|>$ vs. the Sacramento Peak S-index.
}
\label{ff-S}
\end{figure}

The upper panel of Fig.~\ref{ff-S} shows the filling factor ff versus the Sacramento Peak S-index. This figure is comparable to that obtained by \cite{shapiro14}. We also show (lower panel) the average magnetic flux (before the correction described in Sect.~3.1.1) versus the S-index. For a ff of zero, the solar S-index is never below 0.163. However, the minimum flux (basal flux) S-index for solar-type stars is much lower. \cite{mittag13} gave a value of 0.144, which  we consider in the following; this value is compatible with other works \cite[e.g.,][]{gray03,gray06,isaacson10,jenkins11}. This means that for the Sun it is possible to model ff from the S-index or vice versa from the linear law shown in Fig.~\ref{ff-S}, as carried out by \cite{shapiro14} who used it to extrapolate to more active stars.  This law, however, is inadequate to model stars with a lower activity level, since it prevents reaching S-index closer to the basal value; the 0.163 value must therefore be star-dependent. 

We note that for the lowest solar S-index value, the average magnetic field is not zero. This is in part because of the presence of noise, which adds a contribution to all estimates of the average magnetic field and in part because there is a significant amount of magnetic flux outside the
structures defined in Sect.~2.2; the average magnetic fields from the 1 min magnetograms are indeed higher for similar filling factors, while magnetic fields resulting from images averaged over up to one hour are slightly below. 

In the following we therefore assume that the S-index depends on three components as follows: 
\begin{equation}
S(t)=S_{\rm basal}+\Delta S +S_{\rm ff}(t) 
,\end{equation}
where 

$S_{\rm basal}$ is the minimum value of 0.144 obtained by \cite{mittag13}, which is produced when no magnetic field is present.

$S_{\rm ff}(t)$ is proportional to ff according to the law derived from Fig.~\ref{ff-S} and corresponds to strong magnetic fields such as observed in plages and in the magnetic network \cite[variability similar to the one derived by][]{shapiro14}.

$\Delta S$ is 
a necessary component to explain the observation and describes 
the difference between the basal flux and the flux due to network and plage structures. We propose that $\Delta S$ must be due to magnetic fields spread over the rest of the surface, in the so-called quiet Sun. Furthermore, we propose that for a star with a globally lower activity level, for example with cycles lower in amplitude and corresponding to a lower S-index or LogR'$_{\rm HK}$, $\Delta S$ may be lower owing to lower magnetic fluxes over the whole surface. This is the case because less flux comes from the global dynamo either directly or most likely from plages and network, and this lower flux would allow these stars to get a S-index below 0.163 despite a filling factor different from zero. For a given star, $\Delta S$ may vary slightly with time. 
If the magnetic properties in the quiet Sun are constant, the smaller surface coverage of quiet regions at cycle maximum would lead to a slightly smaller $\Delta S$ with a variability proportional to the filling factor of active regions. Conversely, if the magnetic flux in quiet regions is larger at cycle maximum, this should increase $\Delta S$. The resulting variations determine the relative strength of these two terms. 

It is difficult to verify this assumption on the $\Delta S$ component from stellar observations, apart from the fact that low activity stars on average have a smaller variability. However, solar observations may provide clues on the behavior on the weak field regions and $\Delta$S. Results from the literature, which is discussed in Sect.~3.1.3, are indeed not entirely conclusive. We therefore study in more detail the weak field regions of the Sun in the following section, and we focus more specifically on the flux in the most quiet regions over the solar cycle, and on the solar minima conditions.

\section{Solar cycle variations: Quiet Sun component and cycle minimum}

\subsection{Weak field variations over the solar cycle}

\subsubsection{Data and noise analysis}

In this section, we use the same MDI 5 min magnetograms covering a full solar cycle (4181 points). We extracted averages from the absolute value of the magnetic field for various pixel selections in the following section.

Because the noise may impact our results, we implemented the procedure described by \cite{stenflo12} to estimate the noise level on the full disk and for $\mu$ lower than 0.3 for disk-center analysis (as the noise level is lower there). Briefly, the procedure is as follows \cite[we refer to][for more details]{stenflo12}. The method is based on the principle that the noise and magnetic field have different properties as a function of spatial scale. For a given image, we computed the average $<|B|>$ for various resolutions $d$, $B_{\rm obs}(d)$. For a given exponent $k$, we searched for the best true magnetic field $B_{\rm sol}$ (corresponding to no noise at the highest resolution) and noise level $\sigma$ such that $B_{\rm obs}(d)$ after correction from the noise fits $B_{\rm sol}d^{-k}$. When applied to all  magnetograms, this produces two time series: $\sigma(t)$ and $B_{\rm sol}(t)$. We then assumed that the noise level is constant over time, i.e., does not depend on $B_{\rm sol}(t)$. We chose $k$ so that the two time series verify this condition, which in turn gives the noise level $\sigma_0$. We found a slightly lower value of $k$ than the value found by \cite{stenflo12}, i.e., 0.092 instead of 0.13. The difference between the two does not seem to be explained by the formal uncertainties on $k$, which are on the order of 0.005. But we found that subsamples defined by considering the first and second half of the cycle lead to variations up to 0.03, which could in part explain the difference, since the sampling is different. Also, \cite{stenflo12} pointed out the strong sensibility of the $k$ estimation to the procedure, as shown in Hinode data analyzed differently and providing $k$ values differing by a factor 2: \cite{pietarila09} found 0.26 and \cite{stenflo11} found  0.13. A corrected magnetic field $B$ for each pixel of the magnetograms was computed from the measured B$_{\rm app}$, using the following formulae provided by \cite{stenflo12} from \cite{pietarila09} and \cite{stenflo11}: 
\begin{equation}
B=[B_{\rm app}^\alpha-(0.798 \sigma_0)^\alpha]^{[1/\alpha]}
\end{equation}
with
\begin{equation}
\alpha=1.36-0.004\sigma_0+0.0034B_{\rm app}.
\end{equation}
Both B and B$_{\rm app}$ are the magnetic fields divided by $\mu$ before and after correction. 
$\sigma_0$ is the noise level defined by the procedure. 
We found a very similar noise level for the full disk analysis of 19.1~G \cite[][found 18.8~G]{stenflo12}.
We obtained a noise level of 10.8~G for the disk center, where $\mu$ is larger than 0.954, i.e., the relative distance from disk center is lower than 0.3. The advantage of this method is that it is possible to apply a correction, which we did to check for the robustness of our conclusions at the end of next section. We note that when applied, the correction in eq. 4 can only be used for pixels with a large enough magnetic field (above 8.6~G for the 10.8 G noise level for example), therefore the analysis after correction concerns only a small fraction of all available pixels. We also note that the noise estimate is not entirely independent of the activity level. This is not unlike other techniques, such as that used by \cite{jin11}, which consists in fitting the distribution of magnetic fields using a Gaussian. When applied to relative distance from disk center lower than 0.3, we found a similar noise level (values about 1~G below on average) that has a similar variation and a dependence on the cycle as well. This is discussed in Sect.~3.1.2.

\cite{ball12} analyzed MDI magnetograms for the purpose of solar irradiance reconstruuction and found that the SOHO interruption over two periods (June-October 1998 and December 1998-February 1999) led to an increase in noise level mostly in the lower right quadrant of the images. We checked the impact of this effect on our comparison between the solar minima at the end of cycle 22 (i.e., before the interruption) and at the end of solar cycle 23 (i.e., after the interruption) in Sect~3.2.2.

\subsubsection{Average magnetic flux in quiet Sun regions}

\begin{figure*} 
\includegraphics{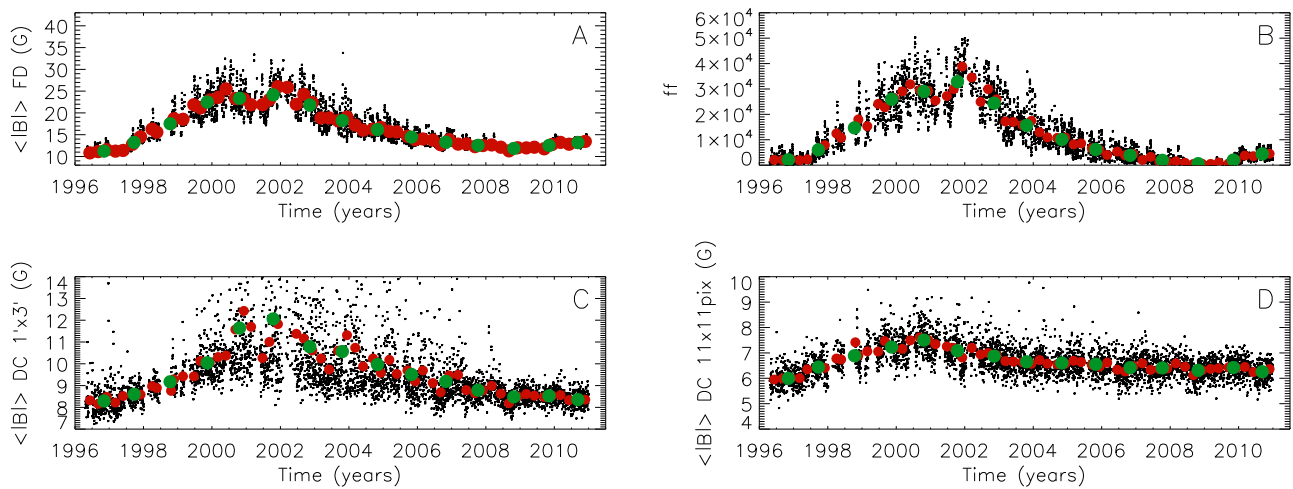}
\caption{
{\it Panel A}: $<|B|>$ averaged over the full disk ($\mu$ lower than 0.95) vs. time from 5 min magnetograms. 
{\it Panel B}: Filling factor ff, corresponding to a 100~G threshold, vs. time. 
{\it Panel C}: $<|B|>$ averaged over the most quiet 1'$\times$3' box close to disk center along the central meridian; a few outliers above 14~G are not shown for clarity.
{\it Panel D}: $<|B|>$ averaged over the most quiet 11$\times$11 pixels box close to disk center along the central meridian.
}
\label{series1}
\end{figure*}

\begin{figure} 
\includegraphics{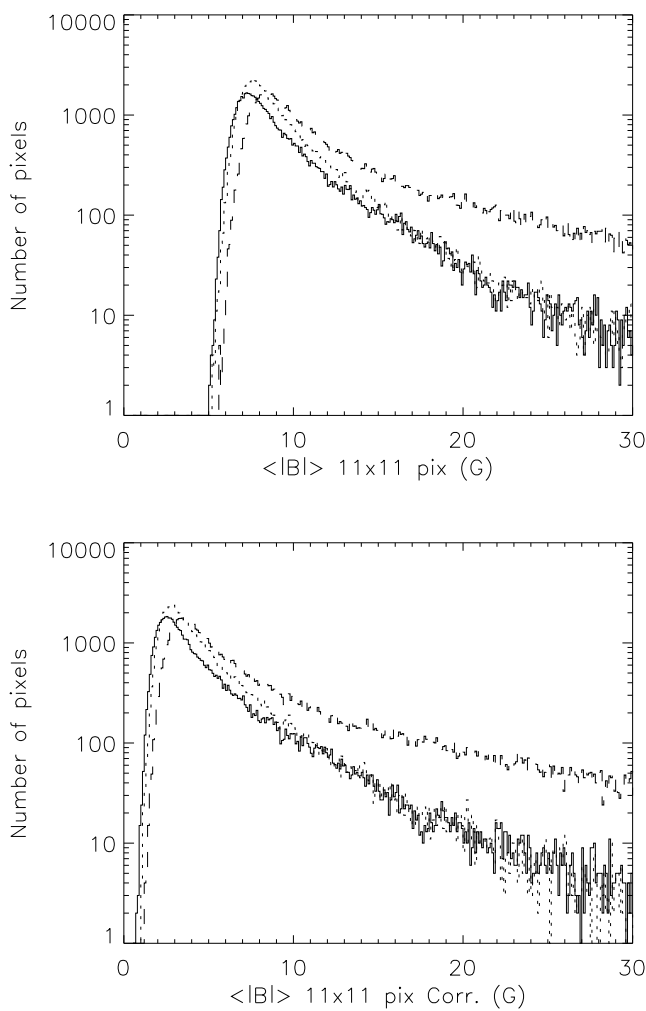}
\caption{
{\it Upper panel}: Distribution of the magnetic field averaged in the 11$\times$11 pixels boxes along the central meridian for three periods: May 1996 - December 1997 (solid line, minimum), January 2000  - December 2003 (dashed line, maximum), and January 2008  - December 2010 (dotted line, minimum).
{\it Lower panel}: Same after correction from the noise level the correction method is described in 3.1.1 and is derived from Stenflo et al. (2012). 
}
\label{histo}
\end{figure}

\begin{figure*} 
\includegraphics{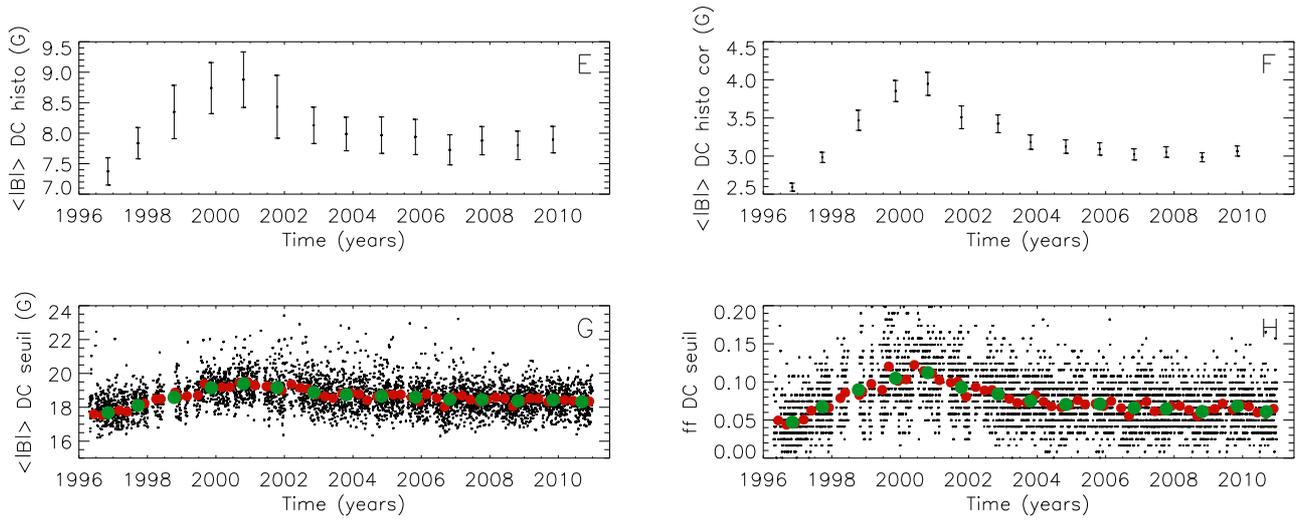}
\caption{
{\it Panel E}: $<|B|>$ derived from the maximum of the distribution of the values along the central meridian (within -7$^{\circ}$ and 7$^{\circ}$ from disk center) in 11$\times$11 pixel boxes.
{\it Panel F}: Same than panel E, but after a correction from the noise. 
{\it Panel G}: Same than panel D, but considering only pixels above 1.5 times the noise level.
{\it Panel H}: Fraction of pixels (in the 11$\times$11 pixel box) used in panel G.
}
\label{series2}
\end{figure*}

\begin{figure} 
\includegraphics{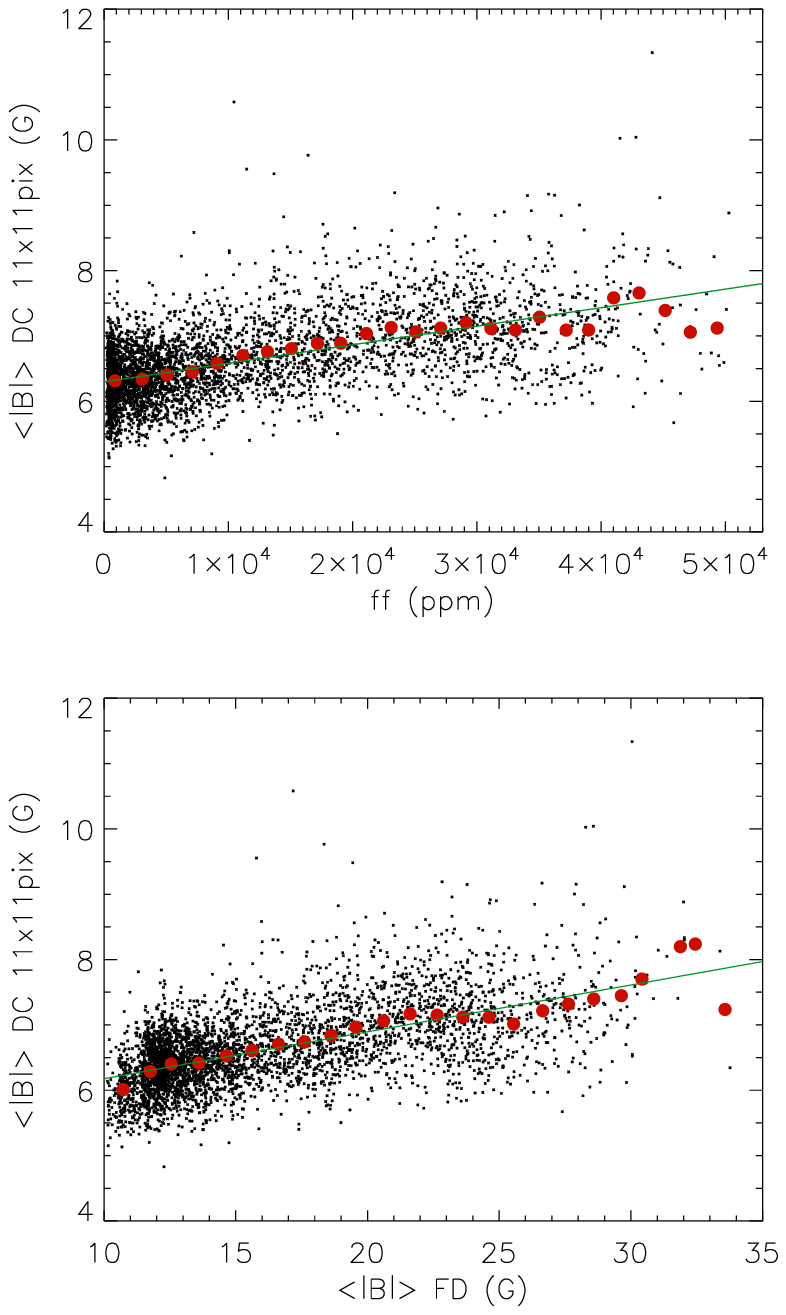}
\caption{
{\it Upper panel}: 11$\times$11 pixel $<|B|>$ vs. ff (ppm) in the quietest area (see text). 
The red dots correspond to a binning of 2000 ppm in ff.
The green solid line is a linear fit on all points.
{\it Lower panel}: Same vs. $<|B|>$ (full-disk) with a bin of 1 G for the red dots.}
\label{dc}
\end{figure}

We first considered the magnetograms without any correction. The two first panels of Fig.~\ref{series1} shows the full-disk $<|B|>$ and ff versus time as derived in the previous section, illustrating the solar cycle variations mostly due to active regions and the magnetic network. 
For comparison, we aimed to estimate the variations in the quiet Sun. For that purpose we defined two indicators and computed the average of |B| over two subsets of pixels.\begin{itemize}
\item{{\it 1'$\times$3' box:} We defined a 1'$\times$3' box on the central meridian between -7$^{\circ}$ and 7$^{\circ}$ in latitude around disk center and localized on the quietest region. A similar procedure (with a visual positioning) was adopted by \cite{livingston07} to search for possible variations of the Ca index at Kitt Peak in the quiet Sun, which allowed  for a comparison. This corresponds to approximately 45$\times$140 Mm, i.e., around nine supergranules. }
\item{{\it 11$\times$11 pixel box:} We defined a 11$\times$11 pixel box using the same procedure. We also studied the distributions of values taken by the average of |B| over such a box along the central meridian and within -7$^{\circ}$ and 7$^{\circ}$. The advantage of this smaller box over the previous example is that a 1'$\times$3' box often contains active network during solar maximum, as already noticed in the Kitt Peak data by \cite{livingston07}. The smaller size used here is equivalent to half a supergranule. Since supergranular cells are outlined by network structures and the interior of the cells correspond to intranetwork field, and given the latitude coverage, this ensured that the minimum values should fall inside a cell without a impact of network features.}
\end{itemize}
The average |B| over each of these boxes is shown versus time on panels C and D of Fig.~\ref{series1}.
In the first case, there are indeed a few outliers that are shown in Fig.~\ref{series1} as a consequence of the presence of magnetic network structures in the box \cite[the same was observed for the chromospheric emission of][]{livingston07}, while such outliers are very rare in the second case. Overall we observed a significant variation over the solar cycle of these two indicators, which has an amplitude on the order of 2 G for the 1'$\times$3' box  and 1.5~G for the 11$\times$11 pixel box;  we computed the value of the former box without the outliers. We note that the results obtained on the noisier 1 min magnetogram time series are similar (on the order  of 1~G). 

The signal in many pixels used in our analysis are below the noise level. 
The signal in such pixels is therefore dominated by the noise, but it also includes some weak true solar signal. The previous analysis therefore took into account all these pixels because they are the majority of pixels and contain some important information. 
We now perform several tests to check the robustness of our results and focus on the 11$\times$11 pixel box analysis. 


We first focus on the distribution of $<|B|>$ values when computed on the 11$\times$11 pixel box, along the central meridian between -7$^{\circ}$ and 7$^{\circ}$ for each image, instead of considering only the minimum value. The distributions of values for three selected temporal ranges are shown in Fig.~\ref{histo} (upper panel); the maximum of the distribution also corresponds to larger magnetic fields during the solar maximum compared to the two solar minima. We note that the values for the 2008-2009 minimum are larger than for the previous minimum; this is discussed in Sect.~3.2. Furthermore, for each 1 y distribution of our time series, we computed the magnetic field corresponding to the maximum of the distribution instead of the average shown in panel D corresponding to the most quiet position. The resulting temporal variation is shown in panel E of Fig.~\ref{series2} and also exhibits a significant variation, correlated with the solar cycle, and has an amplitude on the order  of 1.5~G and a similar shape. We applied the same procedure to the values after correction from the average noise (i.e., on a selection of pixels; see Sect.~3.1.1), and both the distributions (lower panel of Fig.~\ref{histo}) and the temporal variation of the distribution maxima (panel F of Fig.~\ref{series2}) lead to a similar conclusion. On panel F the correction from the noise can be made only for a fraction of the points, and the others are arbitrarily set to zero, which is an underestimation. The actual average magnetic field would therefore lie between the levels shown in panels E and F. 

Finally, we performed another test by computing $<|B|>$ on the same 11$\times$11 box but after eliminating the pixels below 1.5 times the noise level as estimated in Sect.~3.1.1. The results are shown in panel G of Fig.~\ref{series2}, in which we observe a correlation with the solar cycle. We also note that in that case, instead of using all pixels, we consider only 5-15 \% of the pixels depending on the cycle phase. The fraction of used pixels is naturally correlated with the solar cycle, as shown in panel H. 

Because of the dependence of the noise estimate on the activity level (Sect.~3.1.1), the noise estimates cannot be used to distinguish between noise and solar origin for the observed variations. But since it is unlikely that the instrumental noise happens to vary in phase with the cycle, the observed variations are very likely to be of solar origin and variation of the noise estimates owing to  the methods used to measure the noise.  

We conclude that even in the quietest regions of the Sun, the magnetic flux is correlated with the solar cycle, which supports our assumption that $\Delta S$ in stars less active than the Sun on average could be due to a lower flux in quiet regions as well. Fig.~\ref{dc} shows the dependence of the 11$\times$11 pixel $<|B|>$ (panel D of Fig.~\ref{series1}) on the filling factor ff and full-disk $<|B|>$, from which a linear relationship can be extracted. The impact of this result on a possible variation of $\Delta S$ with the average activity level is discussed in Sect.~4.

\subsubsection{Weak magnetic field variations over the cycle}

We now compare the observed variation of the magnetic field in the quietest regions in phase with the solar cycle with those available in the literature.

First of all, our conclusion is compatible with supergranulation properties.   \cite{meunier07} found that the magnetic flux inside supergranular cells is smaller in large cells, while cells are smaller at cycle maximum \cite[][]{singh81, berrilli99, derosa04, meunier08}. This suggests that the magnetic field inside cells is indeed larger at cycle maximum, as we have observed. 


On the other hand, there are a number of indications that below a certain structure flux or size. The number of features are no longer correlated with the solar cycle, but are anticorrelated \cite[e.g.,][]{jin11}. \cite{hagenaar03} also found a lower emergence rate at solar maximum for the smallest ephemeral regions. It is often argued that for that reason there is a possibility that the weakest features are due to a local dynamo process that is different from the global dynamo. However, we propose that such observations are not incompatible with our results. We focus on the estimation of the flux everywhere and not on the number of well-defined structure above the background. In addition, \cite{jin11} found that the distribution of magnetic elements peaks around 10$^{19}$ Mx. A simple shift of the distribution toward larger fluxes per structure at higher activity levels, for example, would then lead to an anticorrelation - correlation pattern similar to the correlation pattern they derive without leading to a smaller background flux. 

\begin{figure} 
\includegraphics{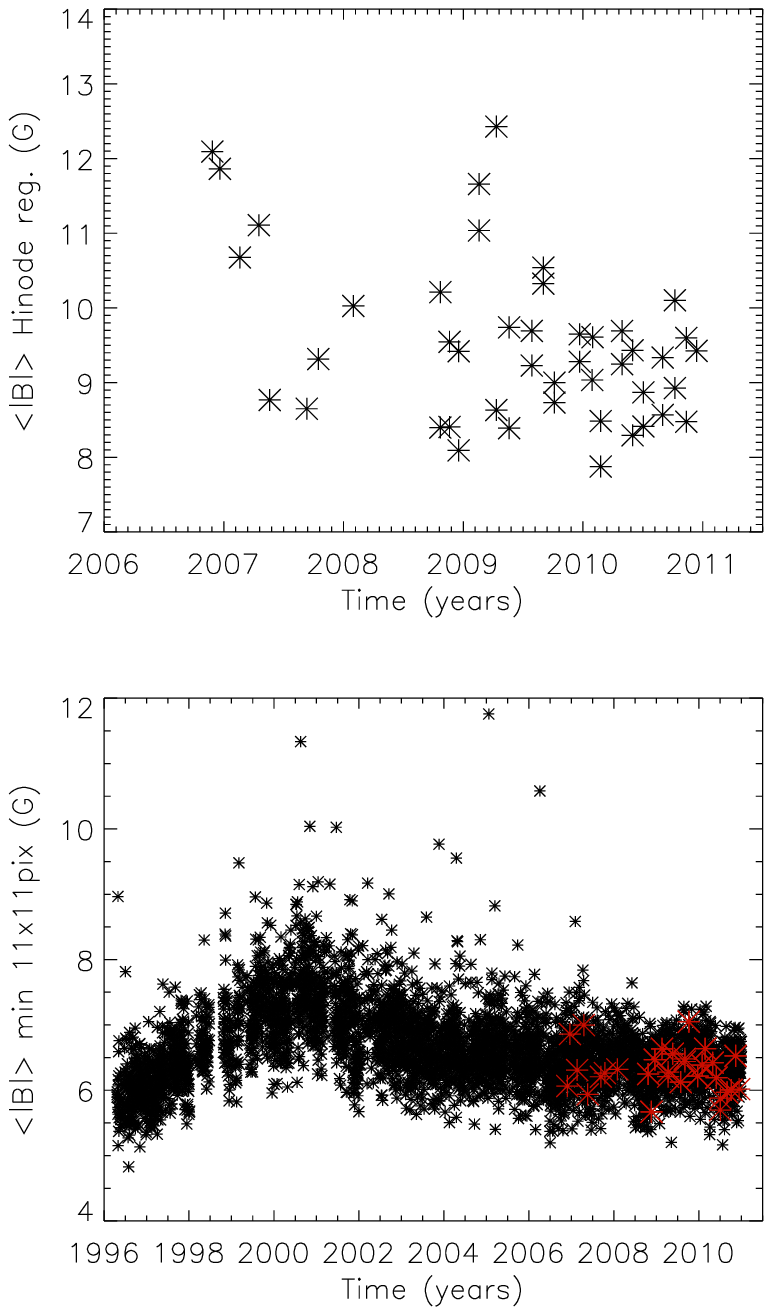}
\caption{
{\it Upper panel}: $<|B|>$ vs. time computed on the same boxes and days than Buehler et al. (2013). 
{\it Lower panel}: 11$\times$11 pixel $<|B|>$ vs. time, for our whole time series (black) and for the days corresponding to the sample of Buehler et al. (2013). 
}
\label{buehler}
\end{figure}

\cite{sanchezalmeida03} defined a few quiet Sun regions on magnetograms during a period of low activity (on a single day of observations), measured the flux for pixels above the noise level, and found that compared to a period of high activity the intranetwork flux did not vary by more than 40\%. This upper limit is therefore not a very strong constraint, which leaves room for a significant variation of the quiet Sun properties during the cycle. For example, panel D corresponds to a 25\% variation, which is compatible with their result.

Our results are also in qualitative agreement with the observations of solar turbulent magnetic fields made by \cite{kleint10} using spectropolarimetric data, showing a long-term variation that is compatible with a weak correlation with the solar cycle. These authors also measured a turbulent field strength around 4.7~G during solar minimum, i.e., between the values derived in panels E and F.

Another similar analysis based on the polarization signal from Hinode observations by \cite{buehler13} led to the conclusion that the magnetic flux in a quiet region close to disk center did not vary over the period November 2006 - May 2012. We rule out that this lack of variation could be because the analysis made by \cite{buehler13} used only a very small part of the signal (about 5\% of the pixels) since, as shown above (panel G of Fig.~\ref{series2}), even with such a selection we find a correlation with the solar cycle.  We propose that the difference between Buehler et al. and our results could be from a sampling effect.  To test whether their results are compatible with ours, we have therefore performed the following analysis. We first computed the average magnetic flux from the 5 min MDI magnetograms for the same box (size, position) and days as \cite{buehler13}. This shows that for this time series of 45 points,  there is indeed no significant variation over the considered time period, as shown in the upper panel of Fig.~\ref{buehler}; the 45 point time series is one point less than their original data (Table 1 in
their paper) because no 5 min magnetogram was available that day. The same is true for the filling factor of structures with |B| above 100 G (not shown here). If we now consider our own criterion (average over 11$\times$11 pixel of $<|B|>$, minimal along the central meridian) over the same temporal sampling (red points on the lower panel in Fig.~\ref{buehler}) the trend is not significant either. But the variation over the whole time series as studied in this paper is significant (black stars). We conclude that if we had performed the analysis presented in Sect.~3.1.2 on a temporal sampling such as that used by \cite{buehler13} we would not have detected a variation either, and we suggest that the lack of variation in their analysis could be due to the much poorer sampling. We note that we have not considered the four points they have included in their plots for the five first months of 2012 because they are not in their table; but given the dispersion in Fig.~\ref{buehler} compared to the timescale of five months, adding such a low number of points over a few months should not change our conclusion.


Finally, \cite{jin15b} concluded that the horizontal fields in the quiet Sun did not vary with the solar cycle, and as they also obtained a constant ratio of the horizontal and vertical fields, this could be incompatible with our results. We could not make the same verification than for the results of \cite{buehler13}, but we suggest that a difference due to the sampling is a possibility as we have not identified any other cause of this difference so far; this should be investigated further in the future. \cite{jin15a} also found that the average magnetic field in structures defined by the pixels above a certain threshold was constant during the solar cycle; This is not necessarily incompatible with our results as there may be more structures above the threshold anyway. \cite{faurobert15} have also analyzed Hinode data and found that the spatial fluctuation in the intranetwork regions were lower at cycle maximum, but do not provide any indication on the possible variations in amplitude. 



\subsection{Solar minima}

In the previous section, we studied the variability of the quiet Sun magnetic field during the solar cycle. We now focus on the conditions during the solar minima and more specifically on the comparison between solar minima. This allows us to discuss the timescales involved.

\subsubsection{Data analysis}

We search for the cycle minima estimated from the MDI time series (filling factor, full disk, and quiet Sun magnetic field in the boxes 1'$\times$3' and 11$\times$11 pixels). We also consider the S-index derived from the Sacramento Peak Observatory, as described in Sect.~2.1. We add the Kitt Peak Ca index whenever available, also converted into a Mount Wilson S-index. The time series are smoothed at different timescales using a running mean: 28 days, 3 months, and 1 year. The minima (date and level) are then estimated from these time series. 

\subsubsection{Comparisons of the last three solar minima}

\begin{table*}
\caption{S-index during cycle minimum}
\label{tabS}
\begin{center}
\renewcommand{\footnoterule}{}  
\begin{tabular}{ll|ll|ll|ll}
\hline
       &  & \multicolumn{2}{c}{End cycle 21} & \multicolumn{2}{c}{End cycle 22} & \multicolumn{2}{c}{End cycle 23} \\ \hline 
Sample & Timescale & S (nb) & t & S (nb) & t & S (nb) & t \\ \hline
Sac. Peak & 1 year &      0.1693 $\pm$  0.0001 (148) &      1985.91 & 0.1680$\pm$  0.0002 (131) &      1996.48 &  0.1679$\pm$  0.0001 (86) &      2007.65 \\
Sac. Peak & 3 months &     0.1686$\pm$  0.0003 (31) &      1985.50 &  0.1675$\pm$  0.0003 (31) &      1996.74 &  0.1661$\pm$  0.0010 (2) &      2008.63 \\
Sac. Peak & 28 d &     0.1679$\pm$  0.0003 (10) &      1985.49 &  0.1663$\pm$  0.0005 (11) &      1996.80 &  0.1654$\pm$      - (1) &      2008.58 \\ \hline
Kitt Peak & 1 year &     0.1687$\pm$  0.0002 (28) &      1986.68 &  0.1672$\pm$  0.0002 (35) &      1996.12 &  0.1671$\pm$  0.0001 (30) &      2009.71 \\
Kitt Peak & 3 months &     0.1680$\pm$  0.0004 (5) &      1986.66 &  0.1662$\pm$  0.0004 (10) &      1996.28 &  0.1662$\pm$  0.0005 (6) &      2008.50 \\
Kitt Peak & 28 d &     0.1671$\pm$  - (1) &      1985.42 & 0.1657$\pm$  0.0004 (7) &      1996.28 &  0.1647$\pm$  - (1) &      2008.57 \\ \hline
\hline
\end{tabular}
\end{center}
\tablefoot{The S-index is shown for each minimum computed at three different timescales from running means. The uncertainties are derived from the dispersion of the values (not computed if only one point in the bin). The number of observing days in the bin is shown between parenthesis. The time t (in years) corresponds to the average over the observing days in the bin identified as the minimum.}
\end{table*}

\begin{table*}
\caption{Magnetic flux and filling factors during cycle minimum}
\label{tabB}
\begin{center}
\renewcommand{\footnoterule}{}  
\begin{tabular}{ll|ll|ll}
\hline
     &  & \multicolumn{2}{c}{End cycle 22} & \multicolumn{2}{c}{End cycle 23} \\ \hline 
Variable & Timescale & S (nb) & t & S (nb) & t \\ \hline
B (FD) & 1 year &    {\bf  11.22}$\pm$  0.04 (316) &      1996.87 & 11.81$\pm$    0.02 (337) &      2008.95\\
B (FD) & 3 months &    {\bf  10.82}$\pm$    0.08 (59) &      1996.47 & 11.27$\pm$    0.05 (86) &      2008.73\\
B (FD) & 28 d &      10.52$\pm$     0.20 (7) &      1996.40 & 10.75$\pm$    0.01 (27) &      2008.74\\
B (CD 1'$\times$3') & 1 year &   {\bf   8.29}$\pm$  0.04 (318) &      1996.91 & 8.44$\pm$    0.02 (338) &      2008.94\\
B (CD 1'$\times$3') & 3 months &  {\bf    8.12}$\pm$   0.05 (84) &      1996.65 & 8.17$\pm$    0.03 (87) &      2008.71\\
B (CD 1'$\times$3') & 28 d &      7.90$\pm$    0.05 (25) &      1997.22 & 7.98$\pm$    0.06 (24) &      2008.77\\
B (CD 11$\times$11pix) & 1 year &   {\bf 6.00} $\pm$ 0.02 (310) &      1996.84 & 6.29$\pm$    0.02 (336) &      2008.98\\
B (CD 11$\times$11pix) & 3 months &  {\bf 5.91} $\pm$ 0.06 (57) &      1996.50 &  6.12$\pm$    0.04 (87) &      2008.68\\
B (CD 11$\times$11pix) & 28 d &     {\bf 5.56} $\pm$ 0.06 (3) &      1996.41 & 5.93$\pm$    0.06 (27) &      2008.74\\
ff (FD) & 1 year &      2103$\pm$      95 (318) &      1996.62 & {\bf 589}$\pm$      20 (352) &      2008.86\\
ff (FD) & 3 months &      1689$\pm$      212 (84) &      1996.74 & {\bf 425}$\pm$      17 (89) &      2008.61\\
ff (FD) & 28 d &      1085$\pm$      226 (26) &      1996.81 & {\bf 346}$\pm$     23 (7) &      2009.64\\
\hline
\end{tabular}
\end{center}
\tablefoot{The average magnetic field at cycle minimum is shown for the full disk (FD), the 1'$\times$3' box in the quietest position at disk center (CD), and for the 11$\times$11 pixels box also in the quietest position at disk center. The filling factor ff (ppm of the solar disk) corresponds to the 100~G threshold. Each of these variables are shown for each minimum computed at three different timescales from the running means. The uncertainties are derived from the dispersion of the values (not computed if only one point in the bin). The number of observing days in the bin is shown between parenthesis. The time t (in years) corresponds to the average over the observing days in the bin identified as the minimum. Values significantly lower than the other minimum at the 1$\sigma$ level are highlighted in bold. }
\end{table*}

We first compare the S-index from Sacramento Peak and Kitt Peak during the 3 minima available in Table~\ref{tabS} on different timescales. The 1985-1986 minimum exhibits a larger Ca emission than the two following minima. However the two minima in 1996 and 2008-2009 are at a similar level with no significant difference. The 2008-2009 is the lowest for the Sacramento Peak index only at the 28 day timescale, but there is only one Kitt Peak point then, therefore this point is less reliable. If we eliminate from the running mean time-series bins with only one point, the Kitt Peak Ca at the 28d scale becomes 0.1675$\pm$0.0019 at the end of cycle 21  and 0.1652$\pm$0.0007 at the end of cycle 23 while the Sacramento Peak at the 28d scale becomes 0.1656$\pm$0.0016 at the end of cycle 23. This does not change our conclusion.

The average magnetic field, either full disk and in the quietest region at disk center, seems to be lower during the 1996 minimum compared to the 2008-2009 minimum, as shown from Fig.~\ref{series1} (panels A, C, and D). This is also illustrated in Table~\ref{tabB}, where the difference is significant when averaged over 1 year and 3 month timescales. It is less significant at the 28 day timescale (except for the 11$\times$11 pixel box). 
As mentioned in Sect.~3.1.1 however, the noise level in the MDI magnetogram has increased, mostly in the lower right quadrant \cite[][]{ball12}, during the SOHO interruptions. Although an increase before and after the two main interruptions (covering June 1998 - February 1999) in our noise estimate is not seen, there is however a small increase at the end of 1997, which does not coincide with the SOHO interruption. The noise estimation at disk center, which should be less sensitive to the behavior of the lower right quadrant \cite[given the maps provided by][]{ball12}, does not show any step during the SOHO interruption. There is a small increase, however, over 1997 mostly. To investigate this further, we performed a similar analysis of the magnetograms by considering separately the upper and lower halves of the images. The lower halves show indeed a larger minimum level at the end of cycle 23, which could be attributed to the noise increase, while the upper halves show a slightly smaller level in 2009 (typically at the 1$\sigma$ level or lower depending on the field of view). It is therefore difficult to detect a significant difference between these two minima from the MDI magnetograms and the magnetic field level is probably very similar during both minima with a possibility for the cycle 23 minima to be slightly lower.

The filling factor (panel C of Fig.~\ref{series1}) exhibits a different behavior; it is significantly lower for the 2008-2009 minimum,  as observed for the spot number (sidc.oma.be). An important difference between the two minima is that during the 1996 minimum we still observe some rotational modulation, while it disappears for a significant length of time (about eight months) during the 2008-2009 minimum, showing that there is indeed a stronger lack of plages and active network then, related to the lower ff. As above, we also computed the filling factor separately for the lower and upper halves of the images. The ratio of the 1996-1997 and 2008-2009 levels is slightly smaller in this case (from 2.5-4.2 for the 1 year timescale down to 1.5-1.7 for the 28 day timescale). However, surprisingly, we would also expect the decrease to be attenuated for the lower parts given that the noise levels increases, but the reverse trend is observed.


\subsubsection{Comparison with previous results}

\cite{livingston10} showed that the Kitt Peak chromospheric emission was not lower in 2008-2009 compared to the previous minimum. We obtained the same results from the Sacramento Peak observation time series. 
\cite{schrijver11} deduced from the Kitt Peak observation that the end of cycle 23 is marginally lower than the previous cycle, which we do not observe from both Kitt Peak and Sacramento Peak indexes unless marginally corresponds to one single observing day, which is not significant. 
The minimum value for the S-index (0.163) is close to that obtained by \cite{baliunas95} or \cite{radick98} from the Mount Wilson survey (0.168). These series are therefore all compatible with each other. 
We note that \cite{schroder12} estimated the S-index from the end of 2008 to mid-2009 to be 0.153$\pm$0.005, which is much lower than previous minima but is not compatible with other observations made at the same time. The uncertainties in calibrations, for example when comparing the Mount Wilson data to others, could explain the difference but only marginally.
However,  the Sacramento Peak data after conversion into an S-index are also compatible with the observations performed by \cite{mittag16} with the TIGRE instrument between August 2013 and May 2015. We conclude that the S-index we used can be compared with the S-index scales to the Mount Wilson values for other stars and lead to compatible results.

Concerning the magnetic field, \cite{schrijver11} argued that the average magnetic field from MDI magnetograms at the end of cycle 23 is well below that of the previous minimum. We find that the difference seems to be small; there is a slightly lower level during the years 2008-2009 compared to 1996 when considering only the upper part of the images. 
\cite{schrijver11} then concluded that the end of cycle 23 was probably representative of a Maunder minimum state, but the analysis of S-index from two independent sources (Kitt Peak and Sacramento Peak) and different criteria from MDI magnetograms analyzed in the present paper contradicts this statement. 
One possibility is the difference in behavior between the filling factor and global magnetic field level could be timescale effects. 
This would be compatible with the results of \cite{mccracken14}, showing that despite the highest level of cosmic rays during the 2008-2009 minimum compared to the previous minimum, the level is still much lower than the level reconstructed for the Maunder minimum, showing that the large-scale heliospheric and solar magnetic fields are far from the Maunder minimum level. These authors also insist on the very long timescale (about 50 years) necessary to reach the extreme levels during the Maunder minimum. This may be compatible with our suggestion that relatively long timescale processes, i.e., much larger than a couple of years in any case, govern the quiet Sun magnetic fields during cycle minima and more generally along the cycle.




To conclude, although a number of authors do not question that the end of cycle 23 was much lower than the previous minimum \cite[e.g.,][]{schrijver11,munoz15}, we find that the relative strength of the two minima strongly depends on the variable considered. It is indeed much lower for a significant length of time in term of plages and network filling factor (as it is for spots), but it is at a similar level for the chromospheric emission (at least within the uncertainties) and similar or slightly lower when considering the magnetic flux over the full disk or in the quietest regions. This could be explained by the impact of the quiet Sun on the global magnetic flux and chromospheric emission, in which the variability is on timescales longer than the duration of the minimum and possibly longer than a solar cycle.


\section{How to derive the solar S-index from filling factors?}

As pointed out in the introduction, stars with a lower average activity level than the Sun tend to have a lower variability as well. Nevertheless, they still exhibit cyclic variation \cite[][]{lovis11b} and rotational modulation \cite[][]{schrijver89,shapiro13}, which suggests that some relatively large features are present. The law relating the network and plage filling factor and a S-index such as that derived by \cite{shapiro14} or as shown in Fig.~\ref{dc} therefore need to be adapted to cover a large range of activity. We propose that it should follow a relationship such as that described in eq.~1, where $\Delta S$ varies very slowly and is strongly dependent on the global activity level of the star on long timescales (possibly several cycles). This is a reasonable assumption because we have shown that the quiet Sun magnetic field indeed varies in phase with the solar cycle, but also probably on longer timescales than the filling factor. The magnetic field in the quietest regions are correlated with the global  activity level (Fig.~\ref{dc}), which suggests that the former is indeed produced by the latter on a relatively long timescale. 
In this section we propose a reconstruction of the solar S-index based on these conclusions.

\subsection{Observational inputs}

A number of studies in the literature provide a relationship between the local magnetic field (at the pixel size) and the chromospheric emission. Such laws can in principle be used to reconstruct a S-index from a magnetogram by integrating them on all pixels. Power laws between the local magnetic field and the chromospheric emission have been derived with two categories of results:

{\it Power law with an exponent in the 0.5-0.7 range}: \cite{schrijver89b} found a power law with an exponent of 0.6 from the study of one large active region. From the analysis of full disk images covering six days and during a period of moderate activity, \cite{ortiz05} found a similar result (exponent of 0.66). The most complete study was performed by \cite{harvey99}, who decomposed the surface into five components corresponding to what these authors call the magnetized atmosphere, i.e., where they could measure a magnetic field. These five components consist of active regions and decaying regions, which have an exponent of 0.47; the enhanced network, which have an exponent of 0.63; and the quiet and weak networks, which have an exponent of 0.47. An important result is that the power laws corresponding to the various categories not only differ by their exponent, but are also significantly shifted, i.e., a plage pixel with a given magnetic field has a larger chromospheric emission than a pixel with the same magnetic field inside a network structure. The two first categories have a relative amplitude of 0.162 and 0.131; the enhanced network has an amplitude of 0.047; and the two last categories, which show a very similar behavior, an amplitude of 0.040. Finally, we note that the relationship derived by \cite{schrijver89b} was qualitatively reproduced by the modelization of chromospheric emission in flux tubes performed by \cite{solanki91}. \cite{cuntz99} also built an atmosphere model to derive the chromospheric flux in magnetic flux tubes for K stars with different rotation rates, which they assumed corresponds to different magnetic fields in flux tubes: the resulting relationship between chromospheric flux and magnetic field is compatible with a power law whose exponent is around 0.5.

{\it Power law with an exponent around 0.3}: \cite{rezaei07} measured the relationship between a chromospheric emission (H-index) and local magnetic fields in relatively quiet regions only, excluding active regions. In the network, their H-index follows a power law with an index of 0.31 (law valid for field above 10~G), while in the intranetwork quiet regions, the relationship seems flat. This concerns magnetic fields below 50~G, most of which are below 10~G, which have a large dispersion in the H-index of about 10\% of the H-index. Thus, these results certainly allow for a small  dependence of the H-index on magnetic field in intranetwork regions, probably up to a variation in their H-index up to 1-2 over the 0-50 G range. \cite{loukitcheva09} found a similar exponent for the network contribution. These two papers, focusing on the network, therefore found a lower exponent than the first category, but do not provide any results for active regions. 
We note that it is not surprising to find a different exponent, and more generally a different behavior for the network and for the intranetwork component, as shown by \cite{rezaei07}. The horizontal component of the magnetic field is important in the weak field regions \cite[e.g.,][]{jin15b}, as opposed to the magnetic network or to plages, so the chromospheric emission could easily be related to the magnetic field in the flux tubes in a different manner. 

\cite{schrijver01} simulated a chromospheric emission from simulated magnetograms and used a power law with a 0.6 exponent, derived from \cite{schrijver89b}, added to a basal flux covering the surface uncorrelated with the magnetic field. 
\cite{schrijver89c} also derived an integrated chromospheric emission from low-resolution synoptic maps and concluded that power laws describing local relationships led to similar power laws describing the so-called flux-flux (between disk-integrated variables) relationships. Conversely, \cite{pevtsov16} used Ca synoptic maps to reconstruct pseudo-magnetograms. 


In the following section, we propose a more complex reconstruction, taking into account the contributions from all magnetic components, most of which are based on the laws provided by \cite{harvey99}. We base this reconstruction on these laws because they are consistent from active regions to the magnetic network and because the various levels corresponding to various kinds of structures are necessary to fit the observations.

\subsection{Reconstruction of a solar S-index: Model}

\begin{figure}
\includegraphics{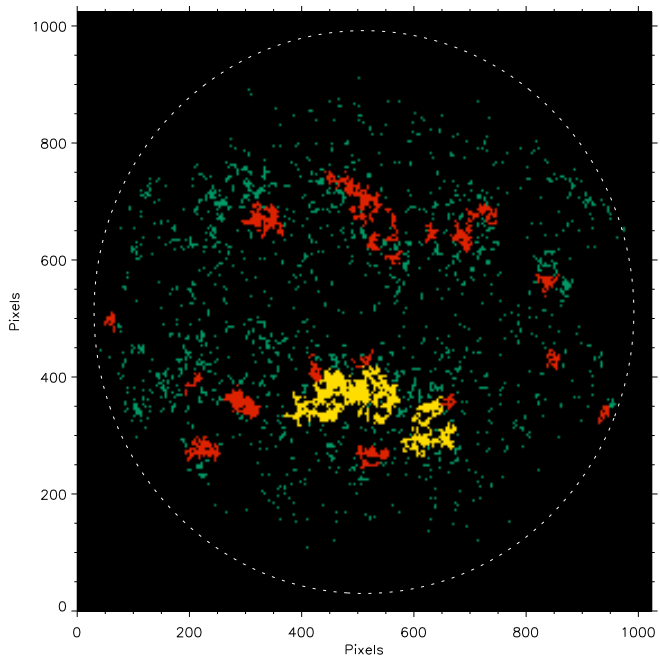}
\caption{
Example of structures identified from an MDI magetogram obtained on 3 June 2002: regions larger 1000 ppm (yellow), regions between 100 and 1000 ppm (red), and structures below 100 ppm (green). The black regions inside the dashed circle (solar limb) correspond to the quiet Sun pixels. 
}
\label{segm}
\end{figure}

\begin{figure} 
\includegraphics{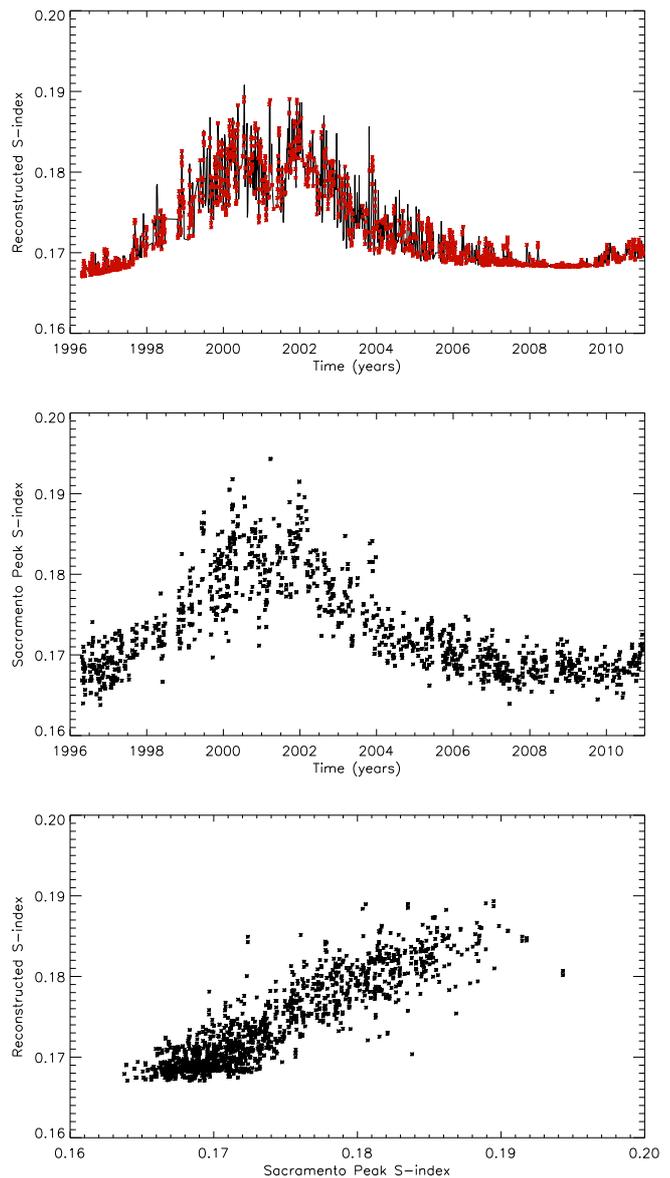}
\caption{
{\it Upper panel}: Example of a reconstructed S-index, for $\beta$=0.3, for all values (black) and those to compare with Ca observations (red).
{\it Middle panel}: Observed Sacramento Peak S-index.
{\it Lower panel}: Reconstructed S-index vs. observed S-index.
}
\label{Srec1}
\end{figure}

\begin{figure} 
\includegraphics{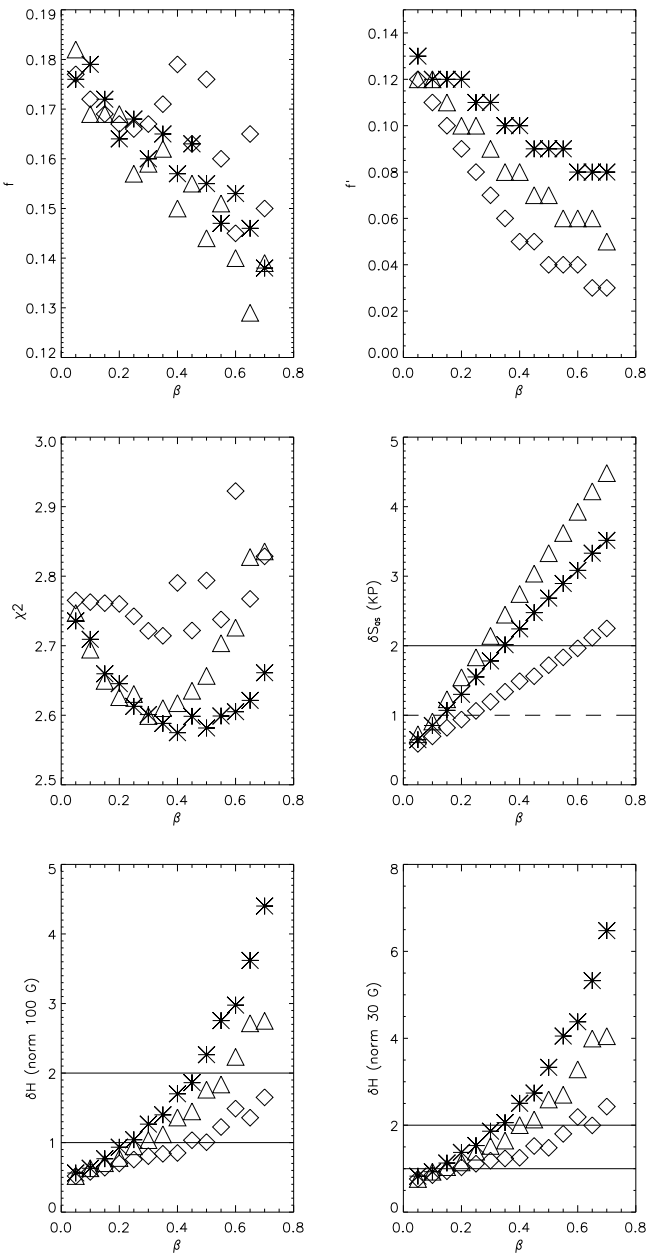}
\caption{
{\it First panel}: f vs. $\beta$ for law 1 (stars), law 2 (diamonds), and law 3 (triangles).
{\it Second panel}: same for f' vs. $\beta$.
{\it Third panel}: Same for chi2 vs. $\beta$.
{\it Fourth panel}: Same for $\delta$S$_{\rm QS}$ in 1'$\times$3' quiet box variability (\%) vs. $\beta$. The horizontal lines highlight the 1 and 2\% levels constrained by Kitt Peak observations.
{\it Fifth panel}: $\delta$H (in units used by Rezaei et al, 2007, normalized to match the results from Harvey et al, 1999, at 100~G)  vs. $\beta$. The horizontal lines highlight H values of 1 and 2, compatible with observations from Rezaei et al (2007).
{\it Sixth panel}: Same for a normalization at 30~G.
}
\label{Srec2}
\end{figure}

In this section we therefore derive a solar S-index from the magnetograms. This approach is in fact similar to that often adopted to reproduce the solar irradiance based on the decomposition of the solar surface in various components contributing to the total solar irradiance, each with its own contrast properties. Our analysis is based on the results obtained in the previous section, suggesting that the field in the quiet Sun is related to the global dynamo. 

We considered three components based on the discussion of eq. 3:

{\it $S_{\rm basal}$ (non magnetic component)}: We considered the value of 0.144 obtained by \cite{mittag13}.

{\it Magnetic component in the active Sun (plage and magnetic network)}: We considered pixels belonging to structures larger than four pixels and defined by a threshold of 40~G to be as close as possible to the conditions of the analysis of \cite{harvey99}. We summed the contribution of all these pixels by applying a slightly simplified version of the laws proposed by \cite{harvey99} to describe only three categories defined by thresholds in size: an average between the active regions and decaying regions law (0.146B$^{0.47}$) for structures largest than 1000 ppm of the solar hemisphere; the enhanced network law (0.047B$^{0.63}$), for structures between 100 and 1000 ppm; the weak and quiet network law (0.040B$^{0.47}$) for the other structures. This segmentation is illustrated in Fig.~\ref{segm}. After division by the total number of pixels, this leads to a proxy for the S-index due to these structures, $S_{\rm act}$, in units similar to that of \cite{harvey99}. We scaled it to the Mount Wilson S-index using a factor $f$, which will be fitted. 

{\it Magnetic component in the quiet Sun}: For the remaining $n$ pixels ($n$ varying over the solar cycle), the magnetograms are significantly  affected by the noise so it is difficult to directly apply a power law to the measured magnetic field to each pixel in observed magnetograms. Instead, we generated at each time step $n$ values of the magnetic field corresponding to the following three laws describing the magnetic field $B_{\rm qs}(t)$ in the quiet Sun: 1) quiet Sun magnetic field from panel F in Fig.~\ref{series2}, which is a lower limit; 2) the average magnetic field over the $n$ pixels in the quiet regions considered, which is an upper limit because it includes the noise (after smoothing at a one year timescale, it varies between 10~G and 12.8~G); and 3) same as the first law, multiplied by an arbitrary factor of two, to study an intermediate situation. We added some dispersion (rms of 2~G) and summed |B|$^\beta$ for these 
pixels, assuming that for intranetwork fields the exponent may be different (possibly smaller) than the exponent for active components: $\beta$ is therefore a free parameter. After division by the total number of pixels, this leads to a $S_{\rm qs}$ in arbitrary units, for which we needed to determine the relative scaling compared to the active contribution, using an unknown constant factor $f$', also a free parameter. This component corresponds to $\Delta S$ in eq. 3, and we expect it to vary much less than $S_{\rm act}$ with time. 

The S-index is then reconstructed as
\begin{equation}
S(t)=S_{\rm basal}+[S_{\rm act}(t)+f' \times S_{\rm qs}(t)]  \times f / N_{\rm pix}(t)
,\end{equation}
where $N_{\rm pix}(t)$ is the total number of pixels in each image, and $f$, $f'$, and $\beta$ are unknown. 
For the 1584 days in common between the MDI series and the Sacramento Peak index, we minimized a $\chi^2$ between this reconstructed S-index and the observed S-index, allowing us to fit $f$ and $f'$ for a given $\beta$, and we considered values of $\beta$ between 0.05 and 0.7. The $\chi^2$ is normalized by $\sigma^2$, where $\sigma$ is an uncertainty on the observed S-index derived by considering the standard deviation of the 56 values between July 2008 and April 2009 (i.e., a value of 0.0014). These data points correspond to a period with a very low and stable filling factor of magnetic structures (almost no rotational modulation), therefore we expect the dispersion in S-index to be mostly due to the uncertainty on the S-index determination.

\subsection{Reconstruction of a solar S-index: Results}

Fig.~\ref{Srec1} shows an example of a reconstructed S-index for $\beta$=0.3 and the first quiet Sun magnetic field law. There is a good agreement with the observed S-index. We note that the dispersion is lower in the reconstructed serie, which is particularly visible at cycle minimum. The reconstructed long-term variation is dominated by the first category of structures (active and decaying regions), which have an an amplitude of 0.015, followed by the second category (enhanced network), which have an amplitude of 0.006. The third contribution is very small and has an amplitude of only 0.0025. The quiet Sun component, which has values between 0.022 and 0.024, as expected varies very little with the solar cycle,  but is comparable in amplitude to the contribution of active regions. 


Fig.~\ref{Srec2} shows the results of the fit as a function of the exponent $\beta$ considered for the quiet Sun component. 
The factors $f$ and $f'$ decrease as $\beta$ increases. The $\chi^2$ does not vary much, but reaches a minimum for $\beta$ in the 0.3-0.5 range. 
The $\chi^2$ is smaller for laws 1 and 3 describing $B_{\rm qs}(t)$, while the upper bound law (law 2) leads to a larger $\chi^2$. 
For the first law (lowest $\chi^2$), the  $\chi^2$ is lowest for $\beta$ in the 0.3-0.5 range, but the variation is small so that other values of $\beta$ are possible, although less likely.  

\subsection{Comparison with observational constraints}

Our objective in this section is to check whether the obtained parameters lead to a chromospheric emission compatible with observational constraints in the quiet Sun. 

\cite{livingston07} showed that the Kitt Peak index measured in 1'$\times$3' regions in quiet regions close to disk center on the central meridian does not exhibit a significant correlation with the solar cycle. These authors estimated that the variation is probably smaller than 1\%, which provides a constraint. After conversion into the S-index this corresponds to a maximum variation of S-index on the order  of 0.0016. We reanalyzed the Livingston et al. data and find that  the dispersion during the two first cycles (21 and 22) of the time series is very large. Nevertheless, the observations performed during cycle 23 (i.e., the cycle we reconstructed)---after binning and taking into account the outliers caused by the passage of small plages or an active network in the box (as on panel C in Fig.~\ref{series1})---show a variation on the order  of 1.5 to 2.2~\% depending on whether the average or the maximum of the distribution is considered. The observations also reveal a shape that is very similar to the average magnetic field in quiet areas versus time, and therefore the Kitt Peak data seems compatible with a variation of the quiet Sun S-index up to $\sim$2\%.

Once we fit the parameters (Sect.~4.1.3), it is possible to reconstruct a S-index following the same recipe (eq. 6) and parameters, but computed only on the 1'$\times$3' box defined in Sect.~3.1.2. We compute the long-term variation from this time series, $\delta$S$_{\rm QS}$.  We then check whether $\delta$S$_{\rm QS}$ is compatible with the 1-2 \% variation from the Kitt Peak constraint, as shown in Fig.~\ref{Srec2}. For laws 1 and 3 describing $B_{\rm qs}(t)$, which are the most probable, $\beta$ should be lower than 0.3 to be compatible with the Kitt Peak observations and large values of $\beta$ can be excluded. Law 2 provides fewer constraints, however. 


A similar conclusion ($\beta$ lower than 0.3 or 0.4 depending on the normalization used) can be drawn from the constraint provided by \cite{rezaei07}, as shown in the lower panels of Fig.~\ref{Srec2}, where we compute the difference in H-index, $\delta$H, between fields of 0 and 50 G and $\delta$H should be lower than 1. We note that to be able to compare our results with those of \cite{rezaei07} described in Sect.~4.1.1, we need to convert our reconstructed S-index into the unit of the H-index studied by \cite{rezaei07}. As their network power law differs from the law obtained by \cite{harvey99}, the normalization depends on the magnetic field considered and is therefore uncertain. We conclude that a $\beta$ exponent to describe the quiet Sun in the range 0.2-0.4 is most probable. 

Finally, \cite{pevtsov13} measured various components of the chromospheric emission, the K2 emission assumed to be representative of active regions, and the K3 emission in the line core and assumed to be associated with quiet regions. The latter also varies during the solar cycle, but it is not clear whether it is contaminated by K2 (with K3 not varying solar cycle in quiet zones) or if quiet zone properties vary. Their results are compatible with a variation of the quiet region chromospheric flux correlated with the solar cycle but it is difficult to use them as a constraint given the possible leakage. We conclude that the obtained range of parameters is compatible with the constraints brought by the observations of the quiet Sun.

\subsection{How to extrapolate to stellar cases?}



The parameters used in the solar S-index reconstruction are not necessarily the same for other stars, and we summarize the important elements below to allow for a reliable stellar S-index reconstruction. 


The parameter $f$, scaling the S-index to the emissions as measured by \cite{harvey99} should not be a critical factor.

The parameter $f'$, scaling the chromospheric emission in quiet area compared to active regions, and the exponent $\beta$, describing the power law relating the quiet region magnetic field to the chromospheric emission could be different in other stars for example although we lack constraints in the literature. \cite{fawzy02c} showed that there may be a trend toward larger chromospheric emission for larger mass stars in the same conditions (magnetic fields, filling factor). 
In active regions, the trend obtained in the models made by  \cite{fawzy02c}  could also be applied. Furthermore, \cite{harvey99} showed that there was a departure from the main power law for low magnetic fields, such that there is a possibility that stars with a lower activity level behave differently (with a weaker chromospheric emission).

The basal flux could be different from the solar flux. We consider the basal flux when no magnetic activity is present by definition, but a possible dependence of this basal flux on spectral type or metallicity could be present. \cite{mittag13} showed that for B-V in the 0.6-0.9 range the basal flux, defined by the lower envelope of the emission versus B-V, should be constant. The theoretical computation of \cite{schroder12} showed that at larger B-V the basal flux increases, but only by a small amount, at least up to B-V around 1.2. The same trend was derived by \cite{fawzy02c}. 
Finally, how the quiet star magnetic field scales with the filling factor, i.e., the total amount of flux in active regions and the active network, is not necessarily proportional. Numerical simulation of magnetic flux tubes in quiet region would help anwser this question. 

\section{Conclusions}

The analysis of a cycle-long time series of MDI magnetograms showed that the magnetic field in the quiet Sun, including in the quietest regions, varies in phase with the solar cycle. This suggests that it is fed by the global dynamo, possibly on relatively long timescales.
The quiet Sun should play a major role in explaining the difference between the basal S-index (corresponding to the most quiet stars) and the S-index at solar minimum. For this reason this property will be very useful in explaining the S-index of stars with a lower activity level on average, as these stars exhibit a lower variability as well, suggesting that they provide less flux for quiet regions. We reconstructed the solar S-index based on properties relating magnetic field and S-index in various types of magnetic regions characterized by \cite{harvey99}. An exponent between 0.2 and 0.4 to describe the power law relating the magnetic field in the quiet Sun and the chromospheric emission is the most likely explanation of the observed solar S-index and is compatible with both the variation of the Ca index in the quiet Sun observed at Kitt Peak by \cite{livingston07} and the  weak dependence of this chromospheric emission on the magnetic field in the intranetwork observed by \cite{rezaei07}. 

\begin{acknowledgements}

This work has been funded by the ANR GIPSE ANR-14-CE33-0018.
This work made use of several public archives and databases: MDI/SOHO archives (SOHO is a mission of international cooperation between the European
Space Agency (ESA) and NASA, ftp://soi-ftp.stanford.edu/pub/magnetograms/);  
Kitt Peak Observatory Ca H and K index (ftp://vso.nso.edu/cycle\_spectra/reduced\_data, NSO-AURA/NSF); 
Ca K index  provided by the Sacramento Peak Observatory of the U.S. Air Force Phillips Laboratory(http://nsosp.nso.edu/cak\_mon)
We are very grateful to the dedication of the observers who have obtained these long-term measurements without which such a work would not have been possible. 

\end{acknowledgements}

\bibliographystyle{aa}
\bibliography{bib30817}

%

\end{document}